\documentclass[11pt]{article}
\usepackage{moriond,epsfig}

\def\be{\begin{equation}}
\def\ee{\end{equation}}
\def\bea{\begin{eqnarray}}
\def\eea{\end{eqnarray}}

\begin{document}
\vspace*{4cm}
\title{Impact of A$_0$ on the mSUGRA parameter space}

\author{Luisa Sabrina Stark$^{1}$, 
	Petra H\"afliger$^{1,2}$, 
	Adrian Biland$^1$ and 
	Felicitas Pauss$^1$}
\address{$^1$Institute for Particle Physics, ETH Z\"urich, CH-8093 Z\"urich, Switzerland \\
	$^2$Paul Scherrer Institut, CH-5232 Villigen PSI, Switzerland}

\maketitle
\abstracts{In mSUGRA models the lightest supersymmetric particle (assumed to be the lightest neutralino $\chi^0_1$) provides an excellent cold dark matter (CDM) candidate. The supersymmetric parameter space is significantly reduced, if the limits on the CDM relic density $\Omega_{CDM}h^2$, obtained from WMAP data, are used.  Assuming a vanishing trilinear scalar coupling $A_0$ and fixed values of tan$\beta$, these limits result in narrow lines of allowed regions in the $m_0 - m_{1/2}$ plane, the so called WMAP strips. In this analysis the trilinear coupling $A_0$ has been varied within $\pm 4$~TeV resulting in largely extended areas in the  $m_0 - m_{1/2}$ plane which are no longer excluded.}

\section{Introduction}
In the mSUGRA framework the lightest neutralino lends itself as an excellent cold dark matter (CDM) candidate, thus providing a connection between particle physics and astrophysics. The inclusion of cosmological experimental data allows to significantly reduce the mSUGRA parameter
space. The satellite born detector WMAP measured the abundance of CDM in the universe to be $0.094\!~<\!~\Omega_{CDM} h^2\!~<\!~0.129$~(at $2\sigma$ C.L.) \cite{WMAP}. \\
The supersymmetric (SUSY) parameter space in mSUGRA scenarios is usually studied in terms of the common scalar mass $m_0$, the common gaugino mass $m_{1/2}$, the ratio of the Higgs expectation
values tan$\beta$ and the sign of the Higgsino mass parameter $\mu$. However, the fifth free parameter, the common trilinear scalar coupling $A_0$, was usually set to zero. In recent studies, the impact of $A_0$ on the SUSY parameter space was recognised \cite{ellisA0}.

\section{Trilinear coupling}
The soft SUSY breaking part of the Lagrangian provides additional contributions to the couplings of Higgs bosons to sfermions. However, as they are proportional to the mass of the corresponding SM fermion, they are only relevant for the third generation. These new couplings affect the masses of the sparticles through renormalisation group evolutions and through mixing effects.
The relic density of the CDM particles is connected to their annihilation cross sections by the Lee-Weinberg equation. A variation of $A_0$ affects the effective cross section and therefore also the relic density through the dependence of the SUSY particle masses on these trilinear scalar couplings at the EW scale. 

\section{Results}
\begin{figure}[htb]
\setlength{\unitlength}{1cm}
\begin{picture}(8,8.2)
\put(-.4,-.2){\epsfig{file=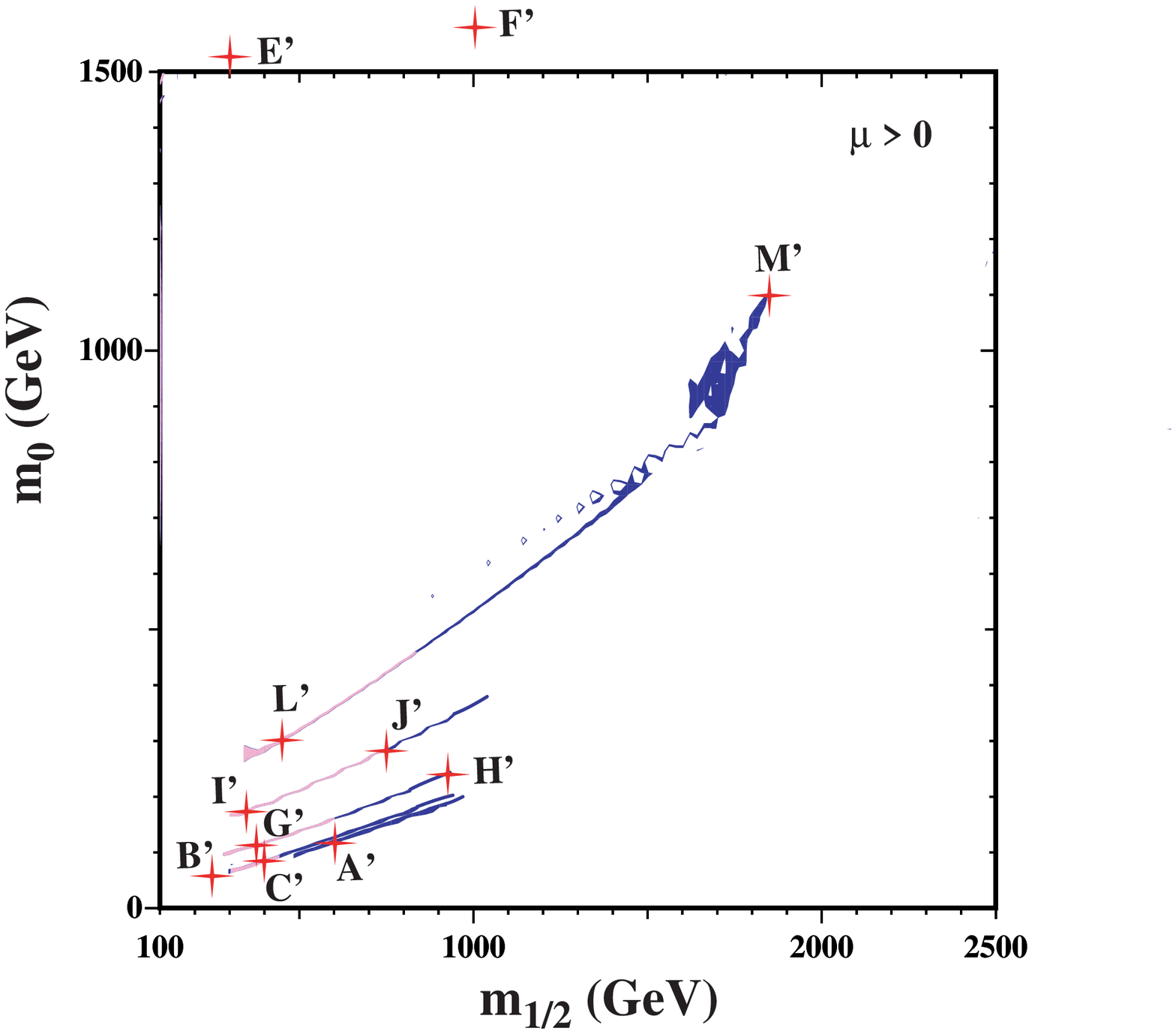,width=9.4cm}}
\put(8.,0){\epsfig{file=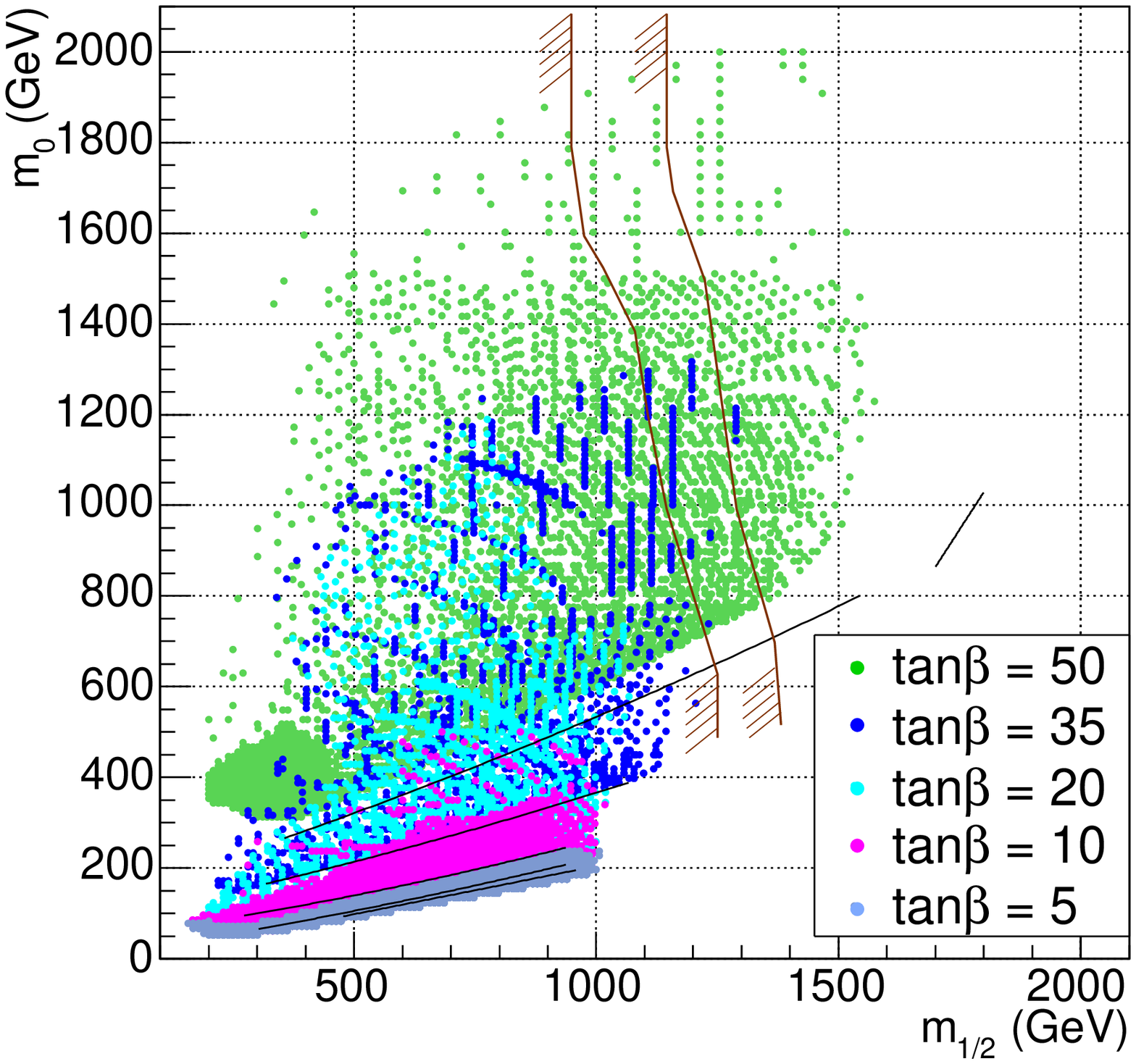,width=8.99cm}}
\put(3.2,4.8){\footnotesize tan$\beta = 50$}
\put(3.2,1.3){\footnotesize tan$\beta = 5$}
\put(1.5,6.99){\footnotesize ($m_t$ = 171 GeV for \bf\scriptsize E', F'\footnotesize)}
\put(4.5,3.2){\footnotesize($m_t$ = 175 GeV)}
\put(9.5,6.9){\footnotesize($m_t$ = 178 GeV)}
\put(14.5,6.9){\footnotesize $\mu>0$}
\end{picture}
\caption{\label{fig1}\it  Allowed models for $m_0~,m_{1/2}\le$~2~TeV, tan$\beta$ between 5 and 50, $\mu~>~0$. For the WMAP strips in the left plot a vanishing trilinear scalar coupling was assumed , while $A_0$ was left free within $\pm$4~TeV in the right one. The brown lines indicate the LHC discovery reach for an integrated luminosity of 100~fb$^{-1}$ and 300~fb$^{-1}$.}
\end{figure}

Assuming CDM to consist exclusively of neutralinos, the cosmological bounds on the neutralino relic density $\Omega_{\chi} h^2$ imply strong constraints on the mSUGRA
parameter space. Under the assumption of $A_0$~=~0 and fixed tan$\beta$, only some narrow lines in the $m_0-m_{1/2}$ plane are left over as allowed regions after including WMAP data \cite{ellis_postW}
(left plot in Fig. \ref{fig1}). In the right plot of Fig.~\ref{fig1} all allowed models for different values of the trilinear scalar coupling $A_0$ and $\tan\beta$ are shown, assuming the Higgsino mass parameter $\mu$ to be positive \cite{A0}. In contrast to the left plot, where only a few lines survived the WMAP constraints, extended regions in the mSUGRA parameter space are allowed, if $A_0$ is varied. For tan$\beta$~=~10 mSUGRA models with $m_0 \sim$~350~GeV for largely negative $A_0$ are within the WMAP constraints, while for $A_0$~=~0 the upper bound was about  $m_0 \sim$~200~GeV. The analogous behaviour can be observed for larger tan$\beta$ values \cite{A0}.

\end{document}